\documentclass[prb,aps,twocolumn,amsmath,amssymb,superscriptaddress,floatfix,longbibliography]{revtex4-2}
\usepackage[breaklinks=true,colorlinks,citecolor=blue,linkcolor=blue,urlcolor=blue]{hyperref}
\usepackage{float}
\usepackage{bm}
\usepackage[normalem]{ulem}
\usepackage{epsfig,mathrsfs,color,latexsym,subfigure,marginnote,graphicx,verbatim,mathrsfs,color,array,amsmath,amsfonts,amssymb,graphicx}
\usepackage{physics}

\newcommand{\rvs}[1]{\textcolor{blue}{#1}}
\usepackage[utf8]{inputenc}
\usepackage{physics}

\begin{document}

\title{High-order van Hove singularities and nematic instability in the kagome superconductor CsTi$_3$Bi$_5$}

\author{Bikash Patra}
\affiliation{Department of Condensed Matter Physics and Materials Science, Tata Institute of Fundamental Research, Mumbai 400005, India}

\author{Amrita Mukherjee}
\affiliation{Department of Condensed Matter Physics and Materials Science, Tata Institute of Fundamental Research, Mumbai 400005, India}

\author{Bahadur Singh}
\email{bahadur.singh@tifr.res.in}
\affiliation{Department of Condensed Matter Physics and Materials Science, Tata Institute of Fundamental Research, Mumbai 400005, India}


\begin{abstract}

ATi$_3$Bi$_5$ (A = Cs or Rb) are emerging topological kagome metals that exhibit superconductivity and nematicity without intertwining translational symmetry-breaking charge orders. In this work, we explore the fermiology of their titanium kagome electrons and identify a set of sublattice-pure, high-order van Hove singularities (VHSs) that can suppress charge ordering and enhance electronic correlations and superconductivity. Our calculations of charge susceptibility for kagome bands with both normal and high-order VHSs emphasize the role of these VHSs in driving electronic nematicity in CsTi$_3$Bi$_5$. Additionally, we compute the phonon spectrum and electron-phonon interactions for CsTi$_3$Bi$_5$ under pristine, doped, and kagome-exposed surface conditions, revealing its robustness against structural instabilities while enhancing the superconducting transition temperature. Our work positions ATi$_3$Bi$_5$ as a key platform for investigating superconductivity and electronic nematicity without translational symmetry-breaking states in kagome metals.
\end{abstract}

\maketitle
\clearpage

\section{Introduction}
Kagome materials, with uniformly interlocked hexagonal and triangular atomic networks, provide a fundamental platform for studying the interplay between frustrated geometry, non-trivial topology, and electronic correlations~\cite{kagome1,  kagome_review3}. Their kagome lattice results in Dirac cones that encode non-trivial topology, and van Hove singularities (VHSs) and flat bands, which drive electronic correlations. These electronic features have been observed in various kagome materials, such as the presence of Weyl fermions in the ferromagnet Co$_3$Sn$_2$S$_2$~\cite{Co3Sn2S2_sci, Co3Sn2S2_nat}, chiral spin textures and a significant anomalous Hall effect in the non-collinear antiferromagnet Mn$_3$X (X = Sn and Ge)~\cite{Mn3X_weyl, Mn3X_anomalous}, Dirac fermions and flat bands in FeSn and CoSn~\cite{FeSn, COSn_2}, coexisting magnetism and superconductivity in Cu$_3$Sn$_2$~\cite{Cu3Sn2}, and charge-density-wave (CDW) in antiferromagnetic FeGe~\cite{FeGe}.  

Among the various known kagome materials, the vanadium-based nonmagnetic kagome superconductors AV$_3$Sb$_5$ (A = K, Rb, or Cs) have realized a wide range of correlated states, including CDW~\cite{AV3Sb5_cdw1, AV3Sb5_cdw2, AV3Sb5_cdw3, AV3Sb5_cdw4}, chiral excitonic order~\cite{AV3Sb5_chiral}, pair density wave~\cite{AV3Sb5_pdw1, AV3Sb5_pdw2}, superconductivity~\cite{AV3Sb5_sc1, AV3Sb5_sc2}, and nematic order~\cite{AV3Sb5_nem}, often found in high-temperature superconductors~\cite{nematic_iron}. The interplay between non-trivial topology and the correlated states in these materials has led to time-reversal symmetry-breaking phenomena~\cite{KV3Sb5_ahe1, CsV3Sb5_ahe2} and the potential emergence of Majorana zero modes~\cite{CsV3Sb5_majorana}. Despite the diversity of quantum states in AV$_3$Sb$_5$, the mechanisms driving these states remain unresolved. This is due in part to the complex interplay of electronic and lattice degrees of freedom, both of which can drive the system into low-energy ordered states. In AV$_3$Sb$_5$, multiple VHSs near the Fermi level, arising from kagome electrons,  create an environment prone to electronic instabilities, while the presence of soft phonon modes points to structural instabilities. For instance, the CDW order may result from electronic nesting at the VHSs or through electron-phonon interactions. Similarly, nematic order, which breaks rotational symmetry, might arise from electronic instabilities or as a result of translational symmetry breaking associated with CDW formation. 
In addition, materials with HfFe$_6$Ge$_6$-type double-kagome lattice have been identified as a promising system to study the interplay between CDW and other electronic orders, as these materials exhibit not only CDW~\cite{Arachchige2022,korshunov2023softening, cao2023competing, tuniz2023dynamics,tan2023abundant} but also nematicity~\cite{jiang2024van}, magnetism~\cite{riberolles2022low, zhang2022electronic}, topological properties~\cite{yin2020quantum, li2021dirac} and superconductivity~\cite{shi2022a}. One notable example, ScV$_6$Sn$_6$, demonstrates the coexistence of CDW and nematicity, with the CDW primarily attributed to the Sc and Sn atoms, while the nematic order is largely associated with the kagome lattice. Despite exhibiting a $\sqrt{3} \times \sqrt{3}$ charge order, ScV$_6$Sn$_6$ maintains intra-unit-cell nematic order, preserving translational symmetry~\cite{jiang2024van}.
Given these complexities, recent research has increasingly focused on exploring alternative kagome materials that are devoid of instabilities stemming from both electronic and phonon degrees of freedom.

Recently, kagome materials based on titanium atomic networks, ATi$_3$Bi$_5$ (A = Rb or Cs), isostructural to AV$_3$Sb$_5$ with Ti and Bi atoms substituting V and Sb, have gained attention~\cite{CsTi3Bi5_exp1, CsTi3Bi5_exp2,yang2023observation,CsTi3Bi5_natphy1, CsTi3Bi5_natphy2,yang2024superconductivity,CsTi3Bi5_vhs}. Spectroscopic studies on ATi$_3$Bi$_5$ revealed canonical kagome electronic features, including a flat band, VHSs, and Dirac nodal lines from kagome bands crossing with trivial bands~\cite{CsTi3Bi5_natphy1, CsTi3Bi5_natphy2,yang2023observation}. The superconducting state of these materials remains contentious, with some reports suggesting a superconducting transition temperature (T$_c$) close to 4.8 K, while others found no superconductivity at ambient conditions~\cite{CsTi3Bi5_sc1, CsTi3Bi5_sc2, CsTi3Bi5_sc3}. First-principles calculations predict that ATi$_3$Bi$_5$ falls within the low electron-phonon coupling regime at the Fermi filling, with a calculated $T_c$ of 0.13 K~\cite{Yi2022large}. Notably, transport experiments reveal no CDW anomalies in contrast to AV$_3$Sb$_5$.  However, ATi$_3$Bi$_5$ exhibits direction-dependent scattering wave vectors~\cite{CsTi3Bi5_natphy1, CsTi3Bi5_natphy2,yang2024superconductivity}, indicating a multi-orbital Fermi surface and the emergence of a pure electronic nematic phase, akin to those in high-temperature superconductors~\cite{Fe_nema1, Fe_nema3}. These findings call for detailed investigations of electronic structure and electron-phonon interactions in ATi$_3$Bi$_5$ to clarify the role of electronic and phonon degrees of freedom in the formation of the various quantum states.

In this paper, we provide a comprehensive analysis of the electronic structure and electron-phonon interactions in CsTi$_3$Bi$_5$ using a uniform first-principles framework. We elucidate the multi-orbital nature of the Fermi surface, highlighting contributions from both Ti $d$ and Bi $p$ states, which hybridize to form multi-orbital Fermi pockets. The kagome bands near the Fermi level arise from Ti $d_{x^2-y^2}/d_{xy}$ states, with flat bands situated below the Fermi level, while VHSs and Dirac bands appear above it. Notably, both the VHSs are high-order and sublattice-pure, differing from those in AV$_3$Sb$_5$. Our detailed charge susceptibility calculations indicate that these high-order VHSs reduce Fermi surface nesting and can drive electronic instabilities. We also demonstrate the stability of CsTi$_3$Bi$_5$ against soft phonon modes and translational symmetry-breaking through analyses of phonon dispersions of pristine, charge-doped, and kagome-exposed surface conditions. Moreover, we explore strategies to enhance the superconducting T$_c$ by tuning the Fermi level to align with flat bands or VHSs via hole and electron doping. Our results highlight that CsTi$_3$Bi$_5$ is a unique kagome material that possesses the necessary properties for realizing electronic instabilities such as nematicity and superconductivity, consistent with experimental findings.

\section{Methodology}\label{methods}

Electronic structure calculations were performed within the density functional theory framework with projector-augmented wave potentials~\cite{DFT_1964, blochl1994projector}, using the Vienna ab initio simulation package (VASP)~\cite{kresse1996efficient, kresse1999from}. The Perdew-Burke-Ernzerhof (PBE) parameterization of the generalized gradient approximation (GGA) was employed to consider exchange-correlation effects~\cite{PBE}. An energy cutoff of 330 eV was set for the plane wave basis set, and a $10 \times 10 \times 8$ $\Gamma$-centered k-point mesh was used for Brillouin zone sampling. Lattice parameters and ionic positions were optimized until the residual forces on each ion were less than $1.0 \times 10^{-3}$ eV/atom. The optimized structural parameters are given in Table~\ref{tab1}. We constructed a material-specific tight-binding model Hamiltonian from atom-centered Wannier functions, considering Ti $s$, $d$, and Bi $p$ states~\cite{wannier90}. Finer $k$-mesh electronic structure calculations and charge-susceptibility analyses were performed using the obtained tight-binding Hamiltonian of CsTi$_3$Bi$_5$.

The superconducting $T_c$ of CsTi$_3$Bi$_5$ was obtained by evaluating the electron-phonon interaction based on the Eliashberg spectral function ($\alpha^2F(\omega)$) with Quantum ESPRESSO (QE)~\cite{QE} and EPW~\cite{EPW} codes. The energy spectrum in QE was calculated with PBE-GGA density functional and norm-conserving PBE pseudopotentials 
with standard accuracy from the PseudoDojo library~\cite{pseudodojo} using the relaxed parameters obtained in VASP. A plane-wave energy cutoff of 60 Ry and a charge density cutoff of 480 Ry were used. The phonon spectrum was obtained through density functional perturbation theory. Notably, we confirmed the consistency of the electron and phonon dispersions calculated in VASP and QE codes and found a good agreement between them (see the Supplemental Materials (SM))~\cite{supp}. For electron-phonon coupling, a 
carefully converged fine 32$\times$32$\times$24 $k$-grid and 16$\times$16$\times$12 $q$-grid were used in the bulk Brillouin zone~\cite{supp}. The effects of additional carriers in CsTi$_3$Bi$_5$ were simulated by adding and removing electrons in the unit cell with a neutralizing uniform background charge. For the calculation of susceptibilities, a fine $200 \times 200$ $k$ mesh was used at temperature $T=0.01$ K.

\section{Results and Discussion}\label{results}

\subsection{Crystal structure and electronic state}

\begin{table}[b!]
\caption{Calculated lattice parameters for CsTi$_3$Bi$_5$ using different exchange-correlation functionals. $a$ and $c$ represent in-plane and out-of-plane lattice constants, respectively. {\it d}$_{Ti-Bi}^\parallel$ and {\it d}$_{Ti-Bi}^\perp$ denote the in-plane and out-of-plane interatomic distance between Ti and Bi atoms.}
\scalebox{1.1}{
\begin{tabular}{c c c c c}
\hline \hline
                                                               &    a (\AA)     &     c (\AA)    &    {\it d}$_{Ti-Bi}^\parallel$ (\AA)    &     {\it d}$_{Ti-Bi}^\perp$ (\AA)   \\  \hline
   PBE                                                     &    5.803        &    10.027        &                2.901                                  &                 2.974                             \\ 
  PBE+vdW                                            &     5.745       &      9.482        &                2.873                                   &                2.957                         \\
  SCAN                                                   &    5.759       &      9.656        &                 2.879                                   &                2.938                          \\
  R2SCAN                                              &    5.785       &      9.787        &                 2.892                                   &                2.953                          \\
  SCAN+vdW                                         &    5.752       &      9.435        &                 2.876                                   &                2.934                          \\  \hline
  Exp.~\cite{CsTi3Bi5_exp1}                  &    5.787       &      9.206        &                 2.893                                   &                2.929                          \\
\hline \hline
\end{tabular}}
\label{tab1}
\end{table}

\begin{figure}[t!]
\centering
\includegraphics[width=0.49\textwidth]{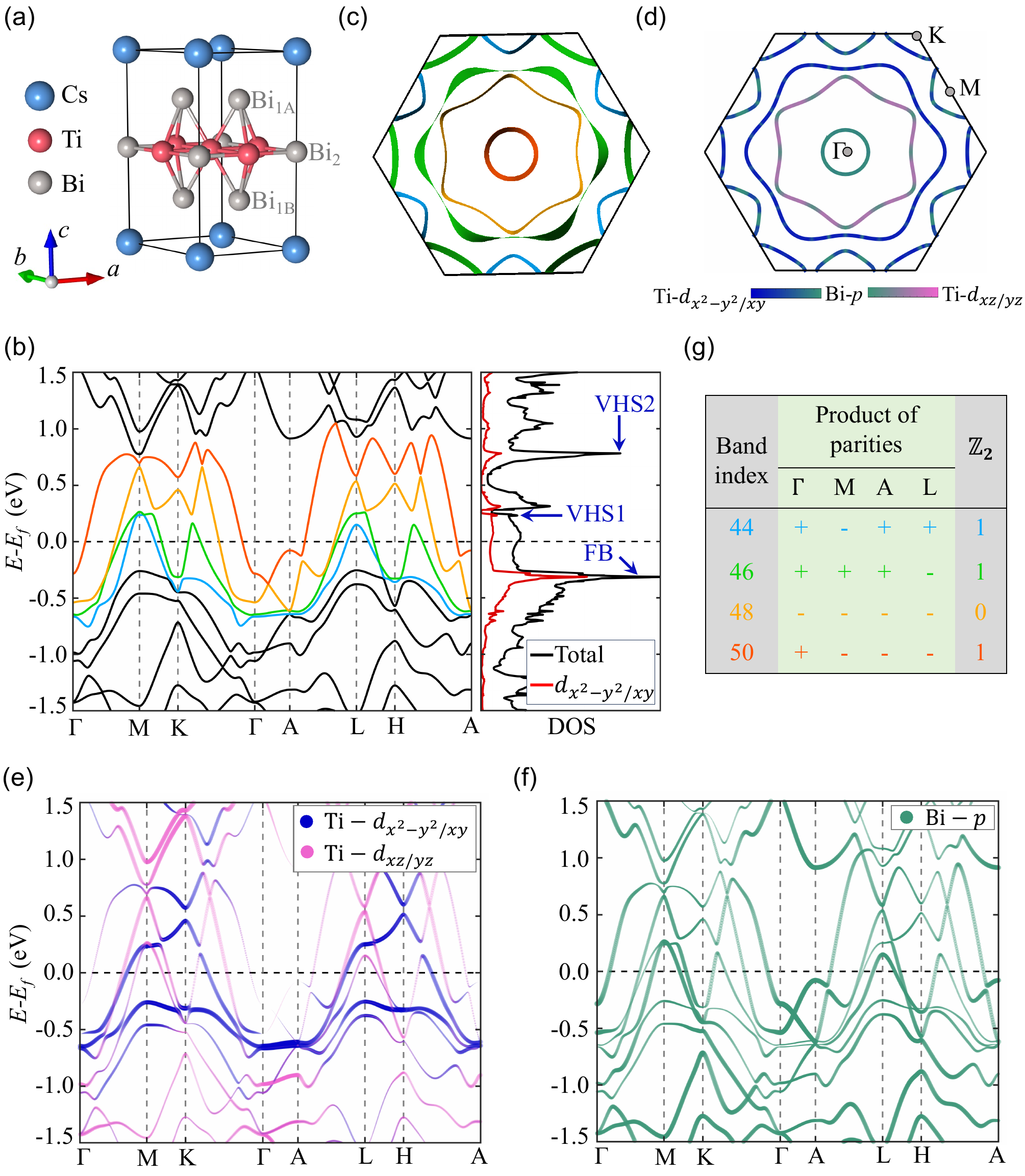}
\caption{(a) Crystal structure of CsTi$_3$Bi$_5$ with Ti kagome net (indicated in red). Bi$_{\text{1A}}$ and Bi$_{\text{1B}}$ represent two honeycomb Bi layers within the lattice. (b) Electronic structure of CsTi$_3$Bi$_5$ with spin-orbit coupling. Blue arrows in the density of states (DOS) highlight saddle-point van Hove singularities (VHS1 and VHS2) and flat band (FB) associated with Ti kagome electrons. (c) Top view of the Fermi surface and (d) orbital-resolved Fermi band contours at the $k_z=0$ plane of CsTi$_3$Bi$_5$.  Two-dimensional Fermi pockets with multi-orbital contributions are evident. (e)-(f) Orbital-resolved band structures of CsTi$_3$Bi$_5$ with (e) Ti and (f) Bi states. Symbol size is proportional to the orbital weight. (g) Parity products at time-reversal invariant momentum points and the associated $\mathbb{Z}_2$ invariant for CsTi$_3$Bi$_5$. } 
\label{fig1}
\end{figure}

CsTi$_3$Bi$_5$ forms a layered hexagonal crystal structure with the space group $P6/mmm$ (No. 191) [Fig.~\ref{fig1}(a)]. In this structure, the Cs and Ti atoms occupy the Wyckoff positions $1a~(0, 0, 0)$ and $3g~(\frac{1}{2}, \frac{1}{2}, \frac{1}{2})$, respectively, while two types of Bi atoms are located at the Wyckoff positions $4h~(\frac{1}{3}, \frac{2}{3}, 0.2387)$ and $1b~(0, 0, \frac{1}{2})$. The structure features a two-dimensional (2D) Ti kagome lattice, with Bi atoms filling the hexagonal centers. This interlaced kagome-Bi lattice is sandwiched between two Bi honeycomb layers that are weakly bonded to the kagome lattice. The Cs atoms form a triangular layer situated between the kagome layers. Table~\ref{tab1} presents the optimized lattice parameters of CsTi$_3$Bi$_5$ obtained using various exchange-correlation functionals. The lattice parameters calculated with the PBE functional are overestimated, particularly the $c$ parameter, which exceeds the experimental value by $\sim8.9$\%. Incorporating van der Waals (vdW) interactions during geometry optimization with PBE reduces the deviation from experimental values. The more advanced SCAN meta-GGA functional, when combined with vdW interactions, yields lattice parameters that closely align with experimental data. Since the lattice parameters derived from PBE and SCAN with vdW interactions differ only slightly, we employ the PBE+vdW optimized parameters to describe the electronic and phononic properties of CsTi$_3$Bi$_5$ in the following discussion.

Figure~\ref{fig1}(b) presents the electronic structure of CsTi$_3$Bi$_5$ with spin-orbit coupling. The band structure features four dispersive band crossings at the Fermi level with similar characteristics at the $k_z=0$ and $k_z=\pi/c$ planes, which highlights the strong 2D nature of the Fermi pockets (Fig.~\ref{fig1}(c)). Despite its complexity, key Kagome features are apparent in the orbital-resolved band structure and partial density of states (Figs.~\ref{fig1}(b) and \ref{fig1}(e)). The kagome bands are predominantly composed of Ti $d_{x^2-y^2}$/$d_{xy}$ orbitals with the saddle-point VHSs (VHS1 and VHS2) and Dirac points located above the Fermi level and flat bands below it.  The Dirac points at the $K$ and $H$ points exhibit small gaps due to spin-orbit coupling. Notably, the VHSs and Dirac points in CsTi$_3$Bi$_5$ are located above the Fermi level, in contrast to those in AV$_3$Sb$_5$ materials, due to the one-electron deficiency of Ti compared to V. 
The position of the flat bands below the Fermi level is ascertained by the orbitals contributing to the kagome bands~\cite{Okamoto2022}. When the kagome bands are formed by $d_{xz}$/$d_{yz}$ orbitals, the flat band lies above the Dirac point. However, when the bands consist primarily of $d_{x^2-y^2}$/$d_{xy}$ orbitals, a sign change in the off-diagonal matrix elements shifts the flat band below the Dirac point, as seen in CsTi$_3$Bi$_5$. 

Figure~\ref{fig1}(c) illustrates the top view of the Fermi surface, comprising five 2D Fermi pockets: three centered at the $\Gamma$ point (one circular, one hexagonal, and one snowflake),  one diamond-like pocket at the $M$ point, and one triangular pocket at the $K$ point. These pockets exhibit substantial hybridization between Ti $d$ and Bi $p$ bands, with varying character across the Fermi surface, as shown in the orbital-resolved Fermi band contours in Fig.~\ref{fig1}(d).  The central circular pocket is derived from Bi $p$ orbitals, while the snowflake pocket is primarily contributed by Ti $d_{x^2-y^2}/d_{xy}$ orbitals that form kagome bands. The remaining three pockets demonstrate strong $d-p$ hybridization, with their character alternating across the Fermi surface, 
resolving a momentum-dependent hybridization (see Figs.~\ref{fig1}(d)-\ref{fig1}(f)).
This multi-orbital Fermi surface is similar to that of the FeSe superconductor, which also exhibits nematic instability. These Fermi surface characteristics are consistent with ARPES studies of CsTi$_3$Bi$_5$, which demonstrate polarization-selective band structure~\cite{CsTi3Bi5_natphy2}. 

We emphasize that, despite the metallic nature of CsTi$_3$Bi$_5$, the band structure exhibits a local energy gap at all time-reversal invariant momentum points. These energy gaps enable the calculations of the $\mathbb{Z}_2$ invariant using the parity criterion~\cite{Fu_Kane}. Figure~\ref{fig1}(g) presents the product of occupied band parities at various momentum points, along with the calculated nontrivial $\mathbb{Z}_2$ invariant. These results indicate that CsTi$_3$Bi$_5$ can host topological surface states within the local energy gaps, as reported in experimental studies~\cite{CsTi3Bi5_natphy2,yang2023observation}.

\subsection{VHSs and nematic instability}
\begin{figure}[t!]
\centering
\includegraphics[width=0.48\textwidth]{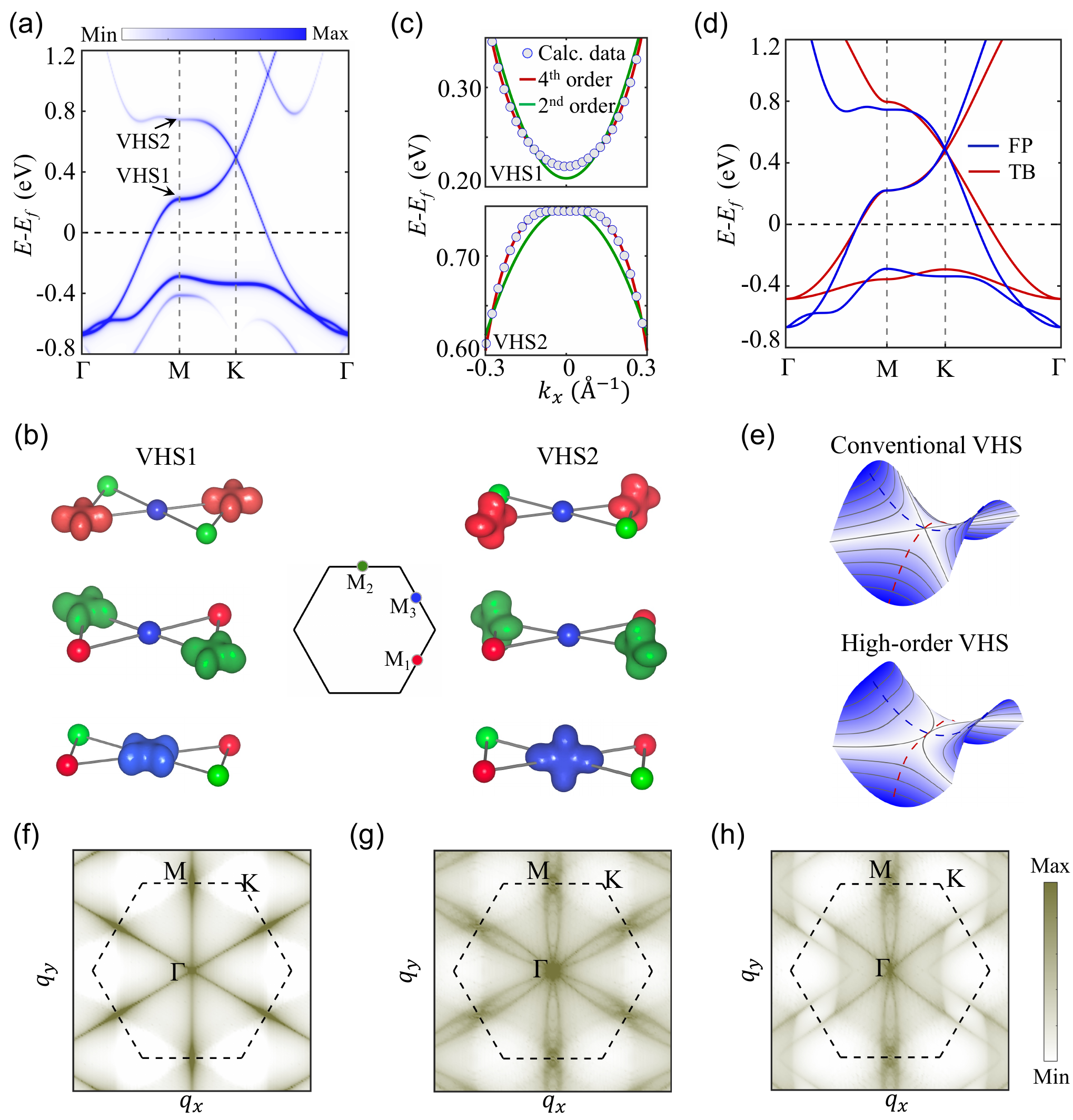}
\caption{(a) Orbital-projected band structure for Ti-$d_{x^2-y^2}/d_{xy}$ states of CsTi$_3$Bi$_5$ at the $k_z=0$ plane, highlighting distinct electronic features of the Ti kagome lattice. (b) Charge density profiles of Bloch wave functions for saddle-point van Hove singularities, VHS1 and VHS2, at three equivalent $M$ points in the Brillouin zone. (c) Energy dispersion along $k_x$ at VHSs, with second and fourth-order polynomial fittings. (d) Calculated band structure of the kagome lattice using the tight-binding model (TB), incorporating nearest and next-nearest neighbor interactions along with first-principles (FP) results. (e) Schematic representation of conventional and high-order VHSs, with grey curves denoting the constant energy contours. (f)-(h) Calculated momentum distribution of bare charge susceptibility $\chi({\bf q})$ for (f) normal VHS (g) high-order VHS derived from CsTi$_3$Bi$_5$ band fittings. (h) Charge susceptibility for high-order VHS by varying onsite energy of one of the kagome sublattices (see text for details).}
\label{fig2}
\end{figure}

We now examine the kagome electronic features and their role in driving electronic nematic order in CsTi$_3$Bi$_5$. Figure~\ref{fig2}(a) presents the Ti $d_{x^2-y^2}$/$d_{xy}$ orbital-resolved band structure, highlighting the Kagome electronic features near the Fermi level. The two VHSs, VHS1 and VHS2, are located at energies 0.22 eV and 0.74 eV above the Fermi level at $M$ point, respectively, while a flat band lies around -0.4 eV below it. These kagome bands cross the Fermi level, forming the Dirac cone at the $K$ point. To characterize the flavor of VHSs, we isolate the sublattice projections of the Bloch wave functions at VHSs, as shown in Fig.~~\ref{fig2}(b). Both VHS1 and VHS2 are sublattice pure, with each sublattice contributing VHS at different M points. Typically, in a kagome band structure, two VHSs exhibit a mix of sublattice pure and mixed characters; however, both VHSs here are sublattice pure without exhibiting mixed characters. This behavior arises from the three-dimensionality of VHS2, resulting from the mixing of Ti $d_{x^2-y^2}$/$d_{xy}$ with Ti $d_{z^2}$ orbitals. The sublattice-pure decoration of VHSs splits the three prominent particle-hole scattering channels in the hexagonal Fermi surface into six, thereby weakening the instabilities like CDW while promoting sublattice interference and long-range interactions~\cite{sublat_interfere_1} 
(see SM of additional details)~\cite{supp}.

To analyze the order of VHSs~\cite{HVHSs_Fu}, we fit the energy dispersion around the saddle points (VHS1 and VHS2) along the $M-K-M$ direction using both quadratic and quartic polynomials, as shown in Fig.~\ref{fig2}(c). The band dispersion is better described by a quartic polynomial $E=a+b k_x^2+ c k_x^4$, with fitting parameters $ a=0.22$, $b=1.07$, and $c=7.63$ for VHS1, and $a=0.74$, $b=-0.33$, and $c=-13.87$ for VHS2. 
The goodness-of-fit parameters, R-square and root mean square error (RMSE), are 0.9999 and 0.0014 for VHS1 and 0.9988 and 0.0003 for VHS2. These values, with R-squared close to 1 and RMSE near 0, confirm the excellent quality of the quartic polynomial fit.
The deviation of energy dispersion from a normal quadratic form indicates the presence of high-order VHSs (Fig.~\ref{fig2}(e)). The high-order VHSs warp the ideal hexagonal Fermi surface associated with normal VHSs, reducing Fermi surface nesting and associated instabilities. Simultaneously, they enhance the DOS, thereby amplifying correlations-driven many-body effects. The sublattice-pure high-order VHSs in CsTi$_3$Bi$_5$ thus suggest a decreased likelihood of CDW instabilities while enhancing sublattice interference and electron correlation effects~\cite{sublat_interfere_2, nematicity_hovhs, wu2021nature}. This interplay can ultimately strengthen electronic instabilities. Such phenomena have been observed in materials like CsV$_3$Sb$_5$~\cite{hovhs_CsV3Sb5} and high $T_c$ cuprates~\cite{hovhs_cuprates}, where significant many-body effects prevail.
  
We now explore how sublattice-pure high-order VHSs weaken CDW instabilities by employing a tight-binding Hamiltonian for kagome bands, 
 \begin{equation}
    H=t \sum_{<i,j>}  c_i^\dagger  c_j  + t^\prime \sum_{<<i,j>>} c_i^\dagger  c_j + H.c.
 \end{equation}
where $t$, and $t^\prime$ are the nearest-neighbour and next-nearest-neighbour hopping parameters, respectively. The energy dispersion for  $t=0.35$ and $t^\prime=0.035$ is shown in Fig.~\ref{fig2}(d). These parameters accurately reproduce the bands at VHS1 but show significant deviations at the VHS2, likely due to contributions from other orbitals and the interference between in-plane and out-of-plane $d$ orbital hoppings. At VHS1, we calculate the bare charge susceptibility $\chi({\bf q})$ using the constant matrix approximation,
\begin{equation}
\chi({\bf q})=-\sum_{{\bf k},m,n} \frac{f(\epsilon_{n{\bf k}})-f(\epsilon_{m{\bf k+q}})}{\epsilon_{n{\bf k}}-\epsilon_{m{\bf k+q}}+i\Gamma}
\end{equation}
where $\epsilon_{m/n,{\bf k}}$ is the eigenvalue of band $m/n$ at ${\bf k}$, $f(\epsilon)$ is the Fermi-Dirac distribution, and ${\bf q}$ is the nesting vector. The susceptibility $\chi({\bf q})$ at normal VHS of a kagome lattice with $t^\prime=0$, shown in Fig.~\ref{fig2}(f), exhibit high intensity along the $\Gamma-M$ lines with peak at the $M$ points. This indicates strong particle-hole scattering linked by nesting vectors ${\bf q}=\{(0,\pi),(\pi,0),(-\pi, \pi)\}$, suggesting formation of potential $2\times 2$ CDW order. However, warped hexagonal Fermi surface due to high-order VHS1 reduces the divergence in $\chi({\bf q})$ at $M$ points (see Fig.~\ref{fig2}(g)). This reduction can be further diminished by considering sublattice decoration to the Fermi surface that modifies the scattering channels, indicating a lower probability of CDW instability. 

The aforementioned discussion indicates the possibility of electronic nematic instability in CsTi$_3$Bi$_5$, as observed in scanning tunneling microscopy measurements~\cite{CsTi3Bi5_natphy1}. Notably, the nematic order in CsTi$_3$Bi$_5$ reduces the $C_6$ rotation symmetry to $C_2$ symmetry. Due to the sublattice pure VHSs, a simple onsite potential can break rotation symmetry, distorting the Fermi surface and lowering the density of states at VHSs~\cite{jiang2024van}. To test this phenomenology, we introduce an onsite potential in one of the sublattices, which effectively modifies the states in one of the three inequivalent $M$ points. The resulting $\chi({\bf q})$, shown in Fig.~\ref{fig2}(h), demonstrates that the $C_6$ symmetry of the system is broken, leading to the emergence of  $C_2$ symmetry. While this model captures the symmetry lowering in the absence of CDW ordering, the multi-orbital Fermi surface and sublattice interference result in non-uniform symmetry breaking, as seen in quasiparticle interference patterns, which warrant further investigation~\cite{CsTi3Bi5_natphy1, CsTi3Bi5_natphy2, yang2024superconductivity}.
    
\subsection{Robustness of structural stability}

\begin{figure}[t!]
\centering
\includegraphics[width=0.48\textwidth]{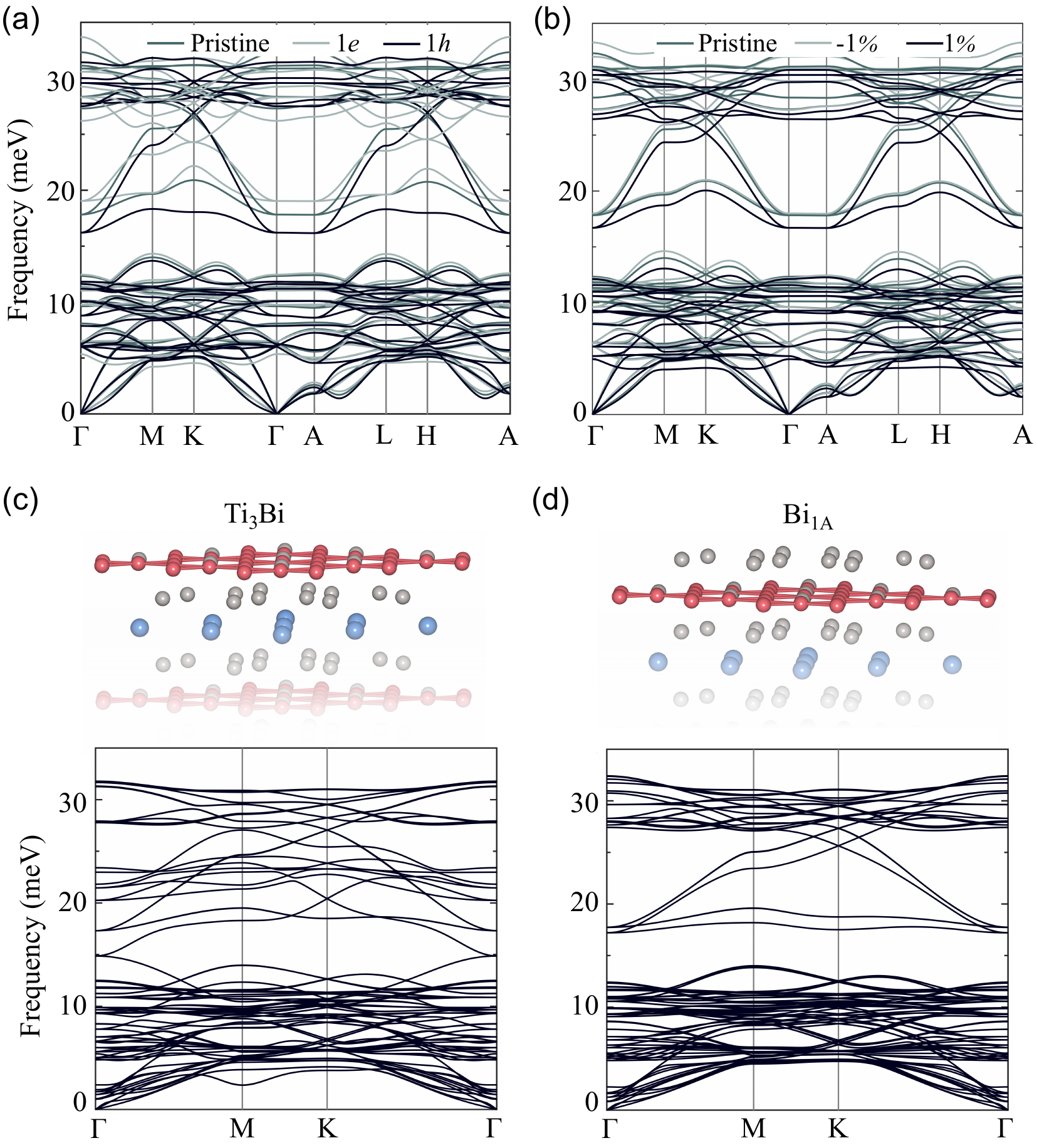}
\caption{(a) Phonon spectrum of CsTi$_3$Bi$_5$ for pristine, electron-doped, and hole-doped phases. (b) Phonon spectrum of CsTi$_3$Bi$_5$ for pristine phase and under compressive and tensile strains. (c)-(d) Phonon spectrum of CsTi$_3$Bi$_5$ slab with (c) Ti$_3$Bi surface termination exposing the Ti-kagome layer and (d) Bi$_{\text{1A}}$ honeycomb layer termination. No imaginary phonon frequencies are evident under these conditions.}
\label{fig3}
\end{figure}

We now examine the possibility of lattice instabilities in CsTi$_3$Bi$_5$. While the VHSs suggest possible electronic instabilities, they are situated slightly above the Fermi level. Therefore, it is essential to investigate dynamical instabilities that may arise from shifting the Fermi level to the VHSs or flat band energies. Figure~\ref{fig3}(a) shows the phonon dispersion for pristine CsTi$_3$Bi$_5$, along with its electron and hole doping cases. In all cases, the phonon eigenvalues remain positive without any imaginary values across the full Brillouin zone. There are only minor changes in the phonon bands even under substantial doping of 1 electron (hole) per unit cell, which moves the Fermi level to the VHS1 (near flat band) energies 
\rvs{(see SM)~\cite{supp}}. 
We also assessed the lattice stability by applying 1\% compressive and tensile strains (Fig.~\ref{fig3}(b)), which similarly yield stable phonon modes. The absence of imaginary phonon frequencies under various conditions suggests that CsTi$_3$Bi$_5$ is robust against lattice instabilities. Notably, even moving the Fermi level to the VHSs or flat bands through doping has no detrimental impact on lattice stability. These results align well with recent experimental reports indicating that CsTi$_3$Bi$_5$ exhibits no structural instabilities even when the VHSs are moved to the Fermi level through Cs doping~\cite{CsTi3Bi5_vhs}.

In bulk kagome materials, the kagome atomic layers are typically buried deep within the structure, complicating the assessment of their influence on structural instabilities. To examine the role of the Ti kagome layer, we calculated the surface phonon dispersions of CsTi$_3$Bi$_5$ for two surface terminations: one that directly exposes the Ti kagome layer and another that positions it just beneath the surface.  The phonon spectra for both terminations, presented in Figs.~\ref{fig3}(c) and \ref{fig3}(d), reveal no imaginary phonon frequencies in the surface Brillouin zone. These results indicate that CsTi$_3$Bi$_5$ retains structural stability even with the kagome layer is exposed at the surface, suggesting that any potential instabilities here would be driven by electronic factors. These characteristics position CsTi$_3$Bi$_5$ as a unique kagome system, especially in contrast to isostructural AV$_3$Sb$_5$ materials and bi-kagome layer systems such as AV$_6$Sn$_6$, which exhibit significant lattice instabilities in their surface-exposed kagome layers~\cite{tan2024competing}.

\subsection{Filling dependent electron-phonon interaction and superconductivity}

\begin{figure}[t!]
\centering
\includegraphics[width=0.48\textwidth]{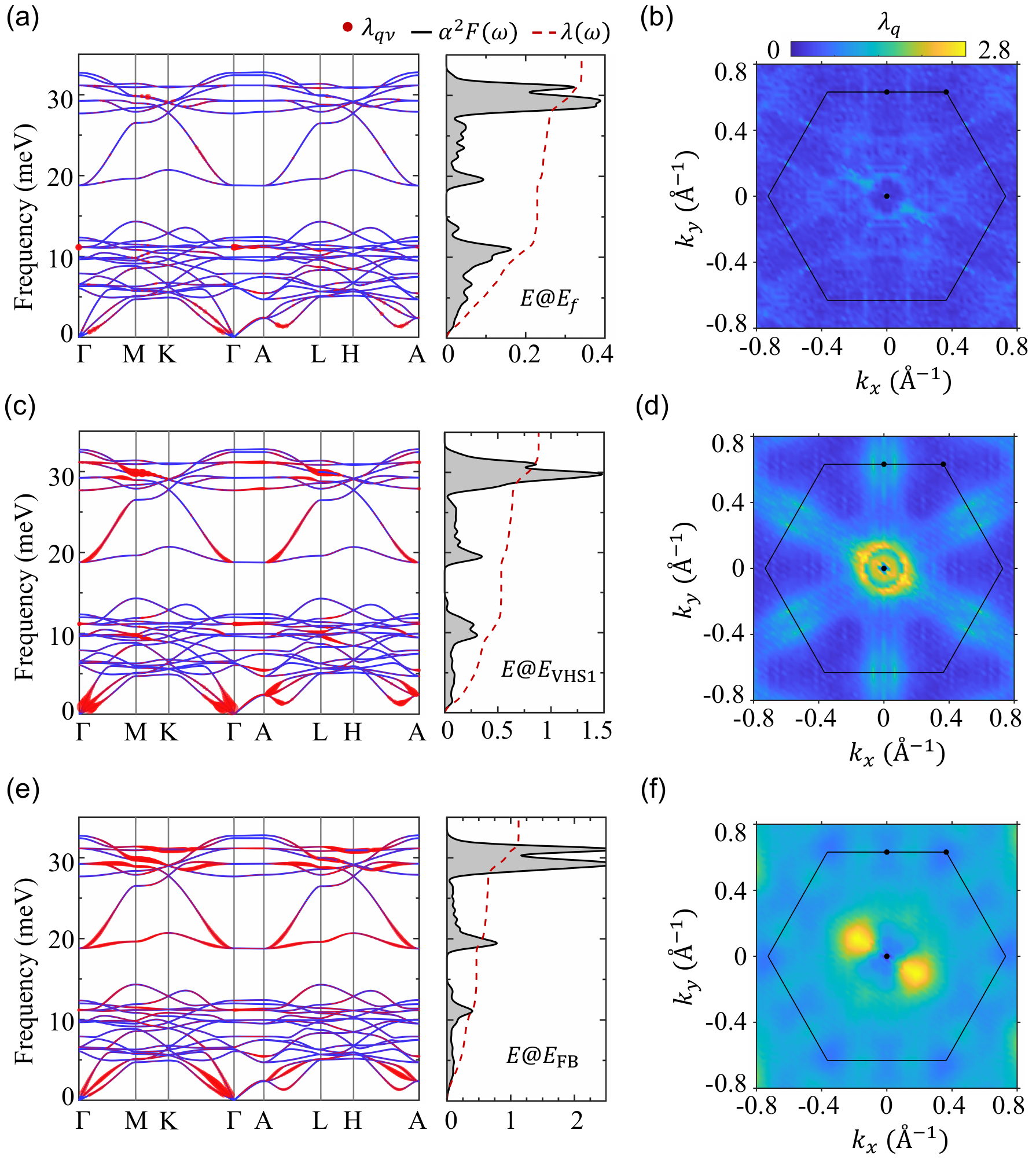}
\caption{(a) Phonon dispersion, Eliashberg spectral function $\alpha^2F(\omega)$ and accumulated electron-phonon coupling constant $\lambda (\omega)$; and (b) distribution of the electron-phonon coupling strength $\lambda_{\nu{\mathbf q}}$ at $k_z=0$  \AA$^{-1}$ plane for CsTi$_3$Bi$_5$ at the charge neutral point. The size of red markers in the phonon spectrum is proportional to the band-resolved electron-phonon coupling strength $\lambda_{\nu{\mathbf q}}$. (c)-(d) and (e)-(f) present the same data as (a)-(b), but obtained for (c)-(d) at VHS filling (electron doping) and (e)-(f) at flat band filling (hole doping). An enhancement of electron-phonon coupling is evident at both VHS and flat band fillings. }
\label{fig4}
\end{figure}

After confirming that CsTi$_3$Bi$_5$ is stable against lattice instabilities, we now explore the possibility of superconductivity in both pristine and doped conditions (electron and hole doping). In Fig.~\ref{fig4}(a), we present phonon spectrum weighted by wave-vector resolved electron-phonon coupling  $\lambda_{\nu{\mathbf q}}$, along with Eliashberg spectral function $\alpha^2F(\omega)$ and the cumulative electron-phonon coupling $\lambda$ at the charge neutrality point. The  Eliashberg spectral function is defined as
\begin{equation}
\small
\alpha^2F(\omega)=\frac{1}{N_f N_{\mathbf{k}} N_{\mathbf{q}}} \sum_{n{\mathbf k},m{\mathbf k}^\prime,\nu}|g^\nu_{n{\mathbf k}, m{\mathbf k}^\prime}|^2     \delta(\epsilon_{n{\mathbf k}})\delta(\epsilon_{m{\mathbf k}^\prime}) \delta(\omega-\omega_{\nu {\mathbf q}})
\end{equation}
where $N_f$ represents the density of states at the Fermi level $\epsilon_f$, $N_{\mathbf{k}}$ and $N_{\mathbf{q}}$ denote the number of $\mathbf{k}$ and $\mathbf{q}$ points. $\epsilon_{n\mathbf{k}}$ and $\omega_{\nu{\mathbf q}}$ give the electronic and phononic eigenvalues and $g^\nu_{n{\mathbf k}, m{\mathbf k}^\prime}$ corresponds to the electron-phonon matrix element. The cumulative electron-phonon coupling $\lambda$ and wave-vector resolved electron-phonon coupling $\lambda_{\nu{\mathbf q}}$ are given by
\begin{equation}
\small
\lambda=\sum_{\nu{\mathbf q}}\lambda_{\nu{\mathbf q}}=2\int \frac{\alpha^2F(\omega)}{\omega}d\omega
\end{equation}

 The Eliashberg spectral function in Fig.~\ref{fig4}(a) reveals four prominent peaks: one in the low-frequency region and the other three in the high-frequency region. This gives a cumulative electron-phonon coupling strength $\lambda=0.33$, indicating a low electron-phonon coupling regime of CsTi$_3$Bi$_5$. While all phonon bands contribute to the total electron-phonon coupling, the contribution of each mode varies significantly. The low-frequency modes account for  $\sim 68$\% of the total electron-phonon coupling (Fig~\ref{fig4}(a)). The distribution of $\lambda_{\nu{\mathbf q}}$ in the $k_z=0$ plane is shown in Fig~\ref{fig4}(b), which reveals substantial variation of $\lambda_{\nu{\mathbf q}}$ in the entire plane.  The associated superconducting critical temperature $T_c$, calculated using the modified McMillan formula~\cite{Allen_Dynes}, 
\begin{equation}
   T_c=\frac{\omega_{log}}{1.2} \exp\Big[ \frac{-1.04(1+\lambda)}{\lambda-\mu^*(1+0.62\lambda)}\Big]
\end{equation}
where $\mu^*$ is the effective screened Coulomb repulsion, and $\omega_{log}=\exp\Big[\frac{2}{\lambda}\int \frac{\alpha^2 F(\omega)}{\omega}log(\omega)d\omega \Big]$ is the logarithmically averaged phonon frequency, yields $T_c \sim  0.13$ K for $\mu^*=0.10$. This very low value of $T_c$ at the charge neutrality point is consistent with experimental results of minimal or absent superconductivity. 

Next, we evaluate the electron-phonon coupling and superconducting $T_c$ under electron doping (1 electron/unit cell), which shifts the Fermi level to VHS1. The band-resolved $\lambda_{\nu{\mathbf q}}$ now spans the entire frequency spectrum while still maintaining its maximum in the low-frequency range (Fig.~\ref{fig4}(c)). This is also reflected in the Eliashberg spectral function, which shows an enhanced peak in the high-frequency region. The high-frequency phonon modes contribute $\sim 38$\% of the total electron-phonon coupling, an increase of 6\% compared to the charge-neutral case. In Fig.~\ref{fig4}(d), we illustrate the momentum-resolved electron-phonon coupling on the $k_z=0$ plane, where the dominant contribution is still centered around $\Gamma$, with additional enhancement near the M points. The calculated $T_c$ rises to 5.66 K. When the Fermi level is tuned to the flat bands by removing two electrons per unit cell (hole doping), $\lambda$ increases to 1.12, resulting in a $T_c$ enhancement to 13.3 K-over 10 times greater than at the charge neutrality point. Such enhancement is expected, as flat bands are known to strengthen electron-phonon coupling and superconducting $T_c$. Given that CsTi$_3$Bi$_5$ is susceptible to charge doping with a tunable band structure~\cite{CsTi3Bi5_vhs}, enhanced superconductivity can be achieved by moving away from the charge neutrality point through both electron and hole doping.

\section{Summary}\label{summary}

We provide the first comprehensive description of the electronic structure and dynamical properties of CsTi$_3$Bi$_5$ through first-principles theoretical modeling. Our analysis reveals the nature of Ti kagome electrons and their role in driving electronic instabilities in CsTi$_3$Bi$_5$. We demonstrate that the Fermi surface exhibits a multi-orbital character, with contributions from both Ti $d$ and Bi $p$ states, which are hybridized throughout the Brillouin zone. The kagome bands near the Fermi level primarily originate from the  Ti-$d_{x^2-y^2}/d_{xy}$ bands and support sublattice pure and high-order VHSs that enhance sublattice interference and long-range electron correlations while reducing Fermi surface nesting. Based on the bare charge susceptibility associated with electrons near the normal and high-order VHSs, we compare and contrast the reduction in particle-hole scattering and demonstrate the emergence of electronic nematic order, characterized by the lowering of $C_6$ rotational symmetry to $C_2$. Our phonon dispersions of pristine, doped, strained, and kagome-layer-exposed surface conditions demonstrate that CsTi$_3$Bi$_5$ is stable against lattice distortions and CDW instabilities. We also explore the enhancement of superconductivity by shifting the Fermi level away from the charge neutrality point, demonstrating a tenfold increase in $T_c$ as the Fermi level approaches the flat band energies. While kagome materials are well known for supporting many-body instabilities, discerning their origins remain challenging. Our results here position CsTi$_3$Bi$_5$ as a unique kagome metal where Fermi surface nesting and soft phonon modes driven CDW instabilities are absent. Instead, the presence of sublattice-pure high-order VHSs and the multi-orbital nature of the Fermi surface clearly signal the presence of electronic instabilities, such as nematicity and superconductivity in agreement with experimental results.

\section*{Acknowledgements} 

We thank S. Mardanya for helpful discussions on susceptibility calculations. This work is supported by the Department of Atomic Energy of the Government of India under Project No. 12-R$\&$D-TFR-5.10-0100 and benefited from the computational resources of TIFR Mumbai. 

\bibliography{CsTi3Bi5}

\end{document}